\documentstyle[12pt,amsfonts]{article}
\parskip=\medskipamount

\def\appendix#1{
  \addtocounter{section}{1}
  \setcounter{equation}{0}
  \renewcommand{\thesection}{\Alph{section}}
 \section*{Appendix \thesection\protect\indent
 \parbox[t]{11.715cm} {#1}}
 \addcontentsline{toc}{section}{Appendix \thesection\ \ \ #1}
  }
\renewcommand{\thefootnote}{\fnsymbol{footnote}}

\newcommand {\cA}{{\cal A}}

\newcommand {\cD}{{\cal D}}

\newcommand {\cN}{{\cal N}}

\newcommand {\cY}{{\cal Y}}

\newcommand{\bA}{{\bf A}}

\newcommand{\bF}{{\bf F}}

\newcommand{\bY}{{\bf Y}}

\def\a{\alpha}
\def \bi{\bibitem}

\def\b{\beta}
\def\c{\chi}
\def\d{\delta}
\def\e{\epsilon}
\def\f{\phi}
\def\g{\gamma}
\def\G{\Gamma}

\def\l{\lambda}
\def\m{\mu}
\def\n{\nu}
\def\o{\omega}
\def\p{\pi}

\def\r{\rho}
\def\s{\sigma}
\def\t{\tau}

\def\x{\xi}
\def\z{\zeta}
\def\D{\Delta}
\def\F{\Phi}
\def\J{\Psi}
\def\L{\Lambda}
\def\O{\Omega}

\newcommand{\ve}{\varepsilon}                            %new
\newcommand{\pa}{\partial}                           %new
\newcommand{\hf}{\frac12}
\newcommand{\vf}{\varphi}
\newcommand{\sect}[1]{\setcounter{equation}{0}\section{#1}}

\newcommand{\be}{\begin{equation}}
\newcommand{\ee}{\end{equation}}
\newcommand{\bea}{\begin{eqnarray}}
\newcommand{\eea}{\end{eqnarray}}
\newcommand{\non}{\nonumber}
\begin{document}
%%%%%%%%%%%%%%%%%%%%%%%%%%
%%%%%%%%%%%%%%%%

\begin{titlepage}
\thispagestyle{empty}

\begin{flushright}
hep-th/0206234 \\
June, 2002
\end{flushright}
\vspace{5mm}

\begin{center}
{\Large \bf On quantum deformation  of conformal symmetry:\\
Gauge dependence via field redefinitions}
\end{center}
\vspace{3mm}

\begin{center}
{\large
S.M. Kuzenko and I.N. McArthur
}\\
\vspace{2mm}

${}$\footnotesize{
{\it
Department of Physics, The University of Western Australia\\
Crawley, W.A. 6009. Australia}
} \\
{\tt  kuzenko@cyllene.uwa.edu.au},~
{\tt mcarthur@physics.uwa.edu.au}
\vspace{2mm}

\end{center}
\vspace{5mm}

\begin{abstract}
\baselineskip=14pt
The effective action in gauge theories is known to depend 
on a choice of gauge fixing conditions. This dependence is 
such that any change of gauge conditions is equivalent to a
field redefinition in the effective action. In this sense, 
the quantum deformation of conformal symmetry in the $\cN=4$ 
super Yang-Mills theory, which was computed in 't Hooft gauge
in hep-th/9808039 and hep-th/0203236, is gauge dependent. 
The deformation is an intrinsic property of the theory in 
that it cannot be eliminated by a local choice of gauge 
(although we sketch a field redefinition induced by a 
nonlocal gauge which, on the Coulomb branch of the theory, 
converts the one-loop quantum-corrected conformal 
transformations to the classical ones). We explicitly 
compute the deformed conformal symmetry in $R_\xi$ gauge. 
The conformal transformation law of the gauge field turns 
out to be $\xi$-independent. We construct the scalar field 
redefinition which relates the 't Hooft and $R_\xi$ gauge 
results. A unique feature of 't Hooft gauge is that it makes 
it possible to consistently truncate the one-loop conformal 
deformation to the terms of first order in derivatives of 
the fields such that the corresponding transformations form 
a field realization of the conformal algebra. 
\end{abstract}

\vfill
\end{titlepage}

\newpage
\setcounter{page}{1}

\renewcommand{\thefootnote}{\arabic{footnote}}
\setcounter{footnote}{0}
%%%%%%%%%%%%%%%%%%%%%%%%%%%
\sect{Introduction}

In gauge theories, not all rigid symmetries of the classical
action can be maintained manifestly throughout 
the quantization procedure, even in the absence of anomalies.
As was demonstrated some years ago by van Holten \cite{vH}
and also discussed in our recent paper \cite{KM},
the problem of maintaining manifestly  a rigid symmetry  
at the quantum level basically reduces to selecting 
covariant gauge fixing conditions. 
The latter cannot always be achieved, 
at least in the class of {\it local} gauge conditions.
A prominent example is provided by conformal 
symmetry (or its supersymmetric extensions)
in quantum Yang-Mills theories with identically 
vanishing beta-function such as the $\cN=4$ super Yang-Mills theory.
It has been known since the early 1970's
(see, e.g.,  \cite{FP1,FP2,TMP} for a more detailed discussion 
and references to the original publications) that, 
if the vacuum in such a theory is conformally
invariant and the gauge field $A_m(x)$ transforms as a primary 
field with the canonical dimension $d_A = 1$, then the quantum theory 
is trivial since the gauge field two-point function is longitudinal, 
$\langle A_m (x_1) \,A_n (x_2) \rangle 
\propto \pa_m \pa_n \ln \,(x_1 - x_2)^2$. 
In other words, the gauge field has no physical 
transverse degrees of freedom, only purely gauge ones.
This clearly demonstrates that no local conformally covariant gauge 
conditions exist\footnote{In the seventies, 
several publications appeared 
(see \cite{BJ} and references therein)
where a nonlocal conformally covariant gauge condition was 
employed for computing some correlation functions in massless QED;
in fact, manifest conformal covariance in this approach was achieved 
by accompanying any special conformal transformation by a gauge one, 
as in \cite{FP1}. Another recipe for achieving a manifest
conformal covariance in massless QED was \cite{BFF}
to use a version of  Gupta-Bleuler 
quantization in conjunction with the higher derivative gauge 
condition $\Box \pa^m A_m =0$ 
(introduced independently in \cite{ES}),
which becomes conformally invariant 
when  the Maxwell equation $\pa^n F_{mn}=0$ is imposed.}. 
As was shown by Fradkin and Palchik \cite{FP1}, 
the generating functional in these theories is invariant under
{\it deformed} special conformal transformations consisting of
a combination of  conformal transformations
and  compensating field-dependent gauge transformations;
the conformal Ward identity associated with the deformed symmetry
leads to a propagator with the correct transverse part.

The approach of \cite{FP1} has recently been 
applied \cite{JKY,KM} to evaluate leading quantum corrections to 
the deformed conformal transformation on the Coulomb branch of the 
$\cN=4$ super Yang-Mills theory. This has led to 
striking results which we summarize here. 
Classically, the action of the $\cN=4$ super Yang-Mills theory 
is invariant under linear conformal transformations which 
in the bosonic sector $\F^i = \{A_m (x) , \,Y_\m (x) \}$, with 
$\m =1,\ldots ,6$, are:
\be 
-\d_{\rm c} A_m = v A_m +\o_m{}^n A_n + \s A_m~, \qquad
-\d_{\rm c} Y_\m =  v Y_\m +\s Y_\m~,
\label{conf}
\ee
where $v= v^m \pa_m$ is an arbitrary conformal Killing vector field,
\be
\pa_m v_n + \pa_n v_m = 2\eta_{mn}\, \s~, \qquad
\s \equiv \frac{1}{4}\pa_m v^m~, \qquad
\o_{mn} \equiv \hf (\pa_m v_n -  \pa_n v_m)~.
\ee
Quantum mechanically, the effective action is invariant 
under conformal transformations which {\it in principle}
receive contributions at each loop order, 
\be
\D \F = \d_{\rm c} \F + \sum_{L=1}^{\infty} \hbar^L \,
\d_{(L)} \F~.
\ee
On the Coulomb branch of the 
$\cN=4$ super Yang-Mills theory, when the gauge group $SU(N+1)$
is spontaneously broken to $SU(N) \times U(1)$, 
the one-loop deformation in the $U(1)$ sector reads
(with $g$ the Yang-Mills coupling constant and $Y^2 = Y_\m Y_\m$)
\be
\d_{(L=1)} A_m = 
- \, \frac{N g^2}{4 \pi^2} \, (\partial^n \sigma)
\frac{F_{mn}}{Y^2}~,
\qquad
\d_{(L=1)} Y_\m = 
\frac{N g^2}{4 \pi^2} \, (\partial^n \sigma)
\frac{\partial_n Y_\m}{Y^2}~,
\label{L=1}
\ee
{\it up to} terms of second order in the derivative expansion.
This deformation was computed in \cite{KM} in the framework 
of the background field approach and with the use of 
't Hooft gauge.
The scalar deformation, $ \d_{(L=1)} Y_\m$, had 
previously been derived in \cite{JKY}.
Modulo a purely gauge contribution, 
the one-loop corrected transformations 
coincide with the rigid symmetry 
\cite{M,Kallosh}
(in what follows, we set $\hbar =1$ 
and introduce $R^4 =N g^2 / (2\pi^2)$)
\bea
\d A_m &=& \d_{\rm c} A_m
-\frac{R^4}{2Y^2}\, (\pa^n \s) \, F_{mn}
+ \pa_m \Big( \frac{R^4}{2 Y^2}\, (\pa^n \s)\, A_n \Big)  ~,
\label{A-def} \\
\d Y_\m &= &\d_{\rm c} Y_\m
+ \frac{R^4}{2 Y^2}\, (\pa^n \s) \, \pa_n Y_\m~,
\label{Y-def}
\eea
of a D3-brane embedded in $AdS_5 \times S^5$ with the 
action (we set $2\p \a' =1$  and ignore the Chern-Simons term,
see, e.g. \cite{Ts} for more detail):
\be
S = - {1 \over g^2} \int {\rm d}^4x
\left( \sqrt{ - {\rm det} \Big( \frac{Y^2}{R^2} \, \eta_{m
n} +  \frac{R^2}{Y^2}\, \partial_{m}Y_\m \partial_{n} Y_\m
+ F_{mn} \Big)} - \frac{Y^4}{R^4}~
\right) ~.
\label{d3brane}
\ee

It was shown by Maldacena \cite{M} that, assuming $SO(6)$
invariance along with 
supersymmetric non-renormalization theorems in 
the $\cN=4$ super Yang-Mills theory \cite{DS}, 
the transformation law (\ref{Y-def}) uniquely fixes 
the scalar part of the D3-brane action (\ref{d3brane}).
${}$From the point of view of  Yang-Mills theory, 
this low energy effective action results from 
summing up quantum corrections to all loop orders.
Thus the one-loop deformation (\ref{Y-def}) of conformal 
symmetry allows us to get non-trivial multi-loop information 
about the effective action! This illustrates that 
the concept of deformed conformal symmetry is clearly 
important and useful. On the other hand, one can 
ask the following natural question: ``Since the deformation 
(\ref{L=1}) corresponds to a particular set of gauge conditions -- 
't Hooft gauge - to what extent is it gauge independent?''
In the present note, we address this question.

This paper is organized as follows.
In section 2, we review some long established results 
(see, e.g., \cite{Lee} and references therein) 
concerning the gauge dependence of the effective action
in gauge theories. 
In particular, we provide a simple proof of the fact that 
any change of gauge fixing conditions
is  equivalent to a field redefinition in the effective action; 
a slightly different proof, based on the use of the BRST symmetry, 
has recently been given in \cite{HM} in the Matrix model context. 
This analysis is extended in section 3 
to the background field quantization scheme, 
which is a convenient way to implement 
a manifestly gauge invariant definition of the effective action. 
In section 4, we specify the sufficient conditions for a rigid
symmetry to become deformed at the quantum level.  
A general discussion of the quantum deformation 
of the conformal symmetry in $\cN=4$ SYM theory
is provided. We also outline the construction 
of a nonlocal field redefinition, 
on the Coulomb branch of $\cN=4$ SYM theory,  
which converts the classical conformal transformation to 
the deformed one.
The gauge dependence of the deformed conformal symmetry 
is analysed in section 5 by explicit calculations in $R_{\xi}$ gauge,
and we make some observations on the significance of 't Hooft gauge.
In section 6, we summarize our results.

\sect{Gauge dependence of the effective action}

We will use DeWitt's condensed notation
\cite{DeWitt67,DeWitt2},
which is by now standard in quantum field theory \cite{Wei};
in particular, $\J_{,i} [\F]$ denotes the variational derivative
$\d \J[\F] /\d \F^i$.
${}$For simplicity, we restrict attention
to the case of bosonic gauge theories.
Let $S[\F]$ be the action of an irreducible gauge theory
(following the terminology of \cite{BV})
describing the dynamics of bosonic fields $\F^i$.
The action is invariant, $S[\F +\d \F] = S[\F]$, 
under gauge transformations
\be
\d \F^i = R^i{}_\a [\F] \, \d\z^\a~,
\ee
with $R^i{}_\a [\F]$ the gauge generators and 
$\d \z^\a$ arbitrary local parameters of compact support.
In what follows, the gauge algebra 
is assumed to be closed,
\be
R^i{}_{\a,j}[\F]\, R^j{}_\b [\F]
- R^i{}_{\b,j}[\F]\, R^j{}_\a [\F]
= R^i{}_\g [\F]\,f^\g{}_{\a \b} [\F]~,
\ee
together with the additional requirements on the gauge generators
\be
R^i{}_{\a,i} [\F] =0~, \qquad f^\b{}_{\a \b} [\F] = 0~,
\label{ad-req}
\ee
which are naturally met in Yang-Mills theories.

Let $\J[\F]$ be a gauge invariant
functional, $\J_{,i} [\F] \, R^i{}_\a [\F]= 0$. 
Under the above assumptions, its chronological vacuum
average $\langle {\rm out} | {\rm T} (\J[\F] )|~{\rm in} \rangle$ 
is known to have a  functional integral representation
of the form
\be
\langle {\rm out} |{\rm T} (\J[\F] )| ~{\rm in} \rangle =
N \int \cD \F \, {\rm Det}  (F[\F]) \, \J[\F]\,
{\rm e}^{ {\rm i} ( S[\F] + S_{\rm GF} [\c [\F ]] )}~,
\label{in-out}
\ee
where $\c^\a [\F]$ are gauge conditions such that the 
${}$Faddeev-Popov operator
\be
F^\a{}_\b [\F] \equiv \c^\a{}_{, i}[\F] R^i{}_\b [\F]
\ee is non-singular. 
The gauge fixing functional $ S_{\rm GF} [\c ]$
is chosen in such a way
that the action $S[\F] + S_{\rm GF} [\c [\F ]] $ is no longer gauge
invariant. In perturbation theory,
it is customary to choose $ S_{\rm GF} [\c ]$
to be of Gaussian form,
$ S_{\rm GF} [\c ] = \hf \c^\a \eta_{\a \b} \c^\b$,
with $\eta_{\a \b}$ a constant non-singular symmetric matrix.

The chronological average 
$\langle {\rm out} | {\rm T} (\J[\F] )|~{\rm in} \rangle$ 
does not depend on the gauge conditions chosen, 
\be
 \langle {\rm out} | {\rm T} (\J[\F] )|~{\rm in} \rangle_{\c + \d \c}
= \langle {\rm out} | {\rm T} (\J[\F] )|~{\rm in} \rangle_\c~,
\ee
with $\d \c^\a [\F] $ a variation of the gauge conditions.
An early proof of this fact \cite{DeWitt67,DeWitt1}
(see also \cite{KM} for a recent review) 
is based on making the change of variables 
\be
\F^i ~\to~ \F^i - R^i{}_\a [\F] \d \z^\a [\F], \qquad
\d \z^\a [\F] = (F^{-1} [\F])^\a{}_\b  \, \d \c^\b [\F]
\label{field-dep}
\ee
in the functional integral 
\be
\langle {\rm out} |{\rm T} (\J[\F] )| ~{\rm in} \rangle_{\c + \d \c}
=
N \int \cD \F \, 
{\rm Det}  (F[\F] +\d F[\F]) \, \J[\F]\, 
{\rm e}^{ {\rm i} ( S[\F] +
S_{\rm GF} [\c [\F ] + \d \c [\F]] )}~,
\ee
with $\d F^\a{}_\b [\F] = \d \c^\a{}_{,i}[\F] R^i{}_\b [\F]$,
and then  using eq. (\ref{ad-req}).

Let $W[J; \c]$ be the generating functional of connected
Green's functions, 
\be
{\rm e}^{{\rm i} W[J;\c]} =
N \int \cD \F \, {\rm Det}  (F[\F]) \,
{\rm e}^{ {\rm i} ( S[\F] + S_{\rm GF} [\c [\F ]]
+ J_i \F^i)}~,
\label{gen-fun}
\ee
and $\G[\f;\c]$ the effective action of the theory, 
\be
\G[\f;\c] = \big( W[J ; \c] -  J_i \, \f^i \big)\big|_{J= J[\f;\c]}~, 
\qquad
\f^i = \frac{ \d}{ \d J_i} W[J;\c]~.
\ee
Both $W[J; \c]$ and $\G[\f;\c]$ depend on the choice of gauge conditions.
This dependence can readily be figured out by making the change
of variables (\ref{field-dep}) in the functional integral 
representation for $W[J; \c +\d \c]$. Then one gets
\be
W[J; \c +\d \c] - W[J; \c ] = -J_i \,
\langle R^i{}_\a [\F] \, (F^{-1} [\F] )^\a{}_\b \,
\d \c^\b[\F ] \rangle ~,
\label{gauge-dep-1}
\ee
where the symbol
$\langle ~~~ \rangle $ denotes
the quantum average in the presence of the source, 
\be
\langle A[ \F ] \rangle
~=~ {\rm e}^{-{\rm i} W[J;\c]} \,
N \int \cD \F \, A[\F] \,{\rm Det}  (F[\F]) \,
{\rm e}^{ {\rm i} ( S[\F] + S_{\rm GF} [\c [\F ]]
+ J_i \F^i)}~.
\ee
Since $\d W /\d\l = \d \G /\d \l$, 
where $\l$ is any parameter in the theory,
and since $J_i = - \d \G /\d \f^i$, 
from eq. (\ref{gauge-dep-1}),
we then derive the following final relation
\be
\G[ \f; \c +\d \c] = \G [\f + \d \f; \c]~, 
\qquad 
\d \f^i [\f;\c]
= \langle R^i{}_\a [\F] \, (F^{-1} [\F] )^\a{}_\b \,
\d \c^\b[\F ] \rangle ~.
\label{gauge-dep-2}
\ee
This relation shows that an infinitesimal change 
of gauge conditions, $\c[\F] ~\to~ \c[\F] + \d \c[\F] $,
is equivalent to a special nonlocal field redefinition, 
$\f^i ~\to~ \f^i +\d \f^i[\f;\c]$, in the effective action.
On the mass shell, $\d \G /\d \f =0$, the effective action is gauge 
independent, and this is known to imply the gauge independence 
of the $S$-matrix (see, e.g. \cite{Lee}).

\sect{Gauge dependence of the effective action \\
in the background field approach}

We now turn to discussing the issue of dependence 
of the effective action on gauge conditions in the framework of
the background field formulation (see
\cite{DeWitt67,tH} and references therein)
which provides 
a manifestly gauge invariant definition of the effective action.
${}$For simplicity, our considerations will be restricted to 
Yang-Mills type theories in which the gauge generators are linear 
functionals of the fields,
\be
R^i{}_{\a,jk} [\F] = 0~.
\ee

In the background field approach, one splits the dynamical 
variables $\F^i$ into the sum of {\it background}
fields $\f^i $ and {\it quantum}
fields $\vf^i$.  The classical action
$S[\f + \vf] $ is then invariant under
{\it background} gauge transformations
\be 
\d \f^i = R^i{}_\a [\f] \, \d \z^\a~, \qquad 
\d \vf^i = R^i{}_{\a,j} \, \vf^j \,\d \z^\a~;
\ee
and {\it quantum} gauge transformations
\be
\d \f^i = 0~, \qquad 
\d \vf^i = R^i{}_\a [\f +\vf] \, \d \z^\a~.
\ee
The background field quantization procedure 
consists of fixing the quantum gauge freedom,
while  keeping  the background gauge invariance intact,
by means of  {\it background  covariant} gauge conditions
$\c^\a [\vf, \f]$.
The effective action is given by the sum of all 1PI Feynman graphs
 which are vacuum with respect to the quantum fields.
Defining the Faddeev-Popov operator 
\be 
F^\a{}_\b [\vf, \f] 
= ( \frac{\d 
}{\d \vf^i}  \, \c^\a [\vf, \f] )
\, R^i{}_\b [\f +\vf]
\ee
and introducing a gauge fixing 
functional $S_{\rm GF} [\c]$, 
which is required to be invariant under 
the background gauge transformations, 
the generating functional of 
connected quantum Green's functions, $ W[J,\f ;\c]$, is given by 
\be
{\rm e}^{{\rm i} W[J, \f ;\c]} =
N \int \cD \vf \, {\rm Det}  (F[\vf, \f]) \,
{\rm e}^{ {\rm i} ( S[\f +\vf ] + S_{\rm GF} [\c [\vf, \f ]]
+ J_i \vf^i)}~.
\label{gen-fun-quan}
\ee
Its Legendre transform
\be
\G[\langle \vf \rangle, \f ;\c] 
= W[J , \f ; \c] -  J_i \, \langle \vf^i \rangle~, 
\qquad
\langle \vf^i \rangle = \frac{ \d}{ \d J_i} W[ J, \f;\c]
\ee
is related to the 
effective action  ${\bf \G} [\f;\c]$ as follows: 
${\bf \G}[\f;\c] =\G[ \langle \vf \rangle =0, \f;\c]$. 
In other words, ${\bf \G} [\f;\c]$ coincides with 
$W[ J,\f ; \c]$ at its stationary point $J = J[\f;\c]$
such that $\d W[ J ,\f; \c] /\d J = 0$.
By construction, ${\bf \G} [\f;\c]$ is invariant 
under the background gauge transformations.

To determine the dependence of ${\bf \G} [\f;\c]$ 
on $\c$, one can start with the functional integral 
representation (\ref{gen-fun-quan}) for 
$W[J, \f;\c +\d \c]$, where $\d \c^\a [\vf, \f]$ 
is an infinitesimal change of the gauge conditions, 
and make in the integral the following replacement  of variables: 
\be
\vf^i ~\to~ \vf^i - R^i{}_\a [\f +\vf] \,
(F^{-1} [\vf,\f])^\a{}_\b  \, \d \c^\b [\vf, \f]~.
\ee
This leads to 
\be 
{\bf \G}[\f;\c+\d \c] - {\bf \G}[\f;\c]
= 
\langle R^i{}_\a [\f +\vf] \,
(F^{-1} [\vf,\f])^\a{}_\b  \, \d \c^\b [\vf, \f] \rangle 
\;
\frac{\d \G[\langle \vf \rangle, \f ;\c]} 
{\d \langle \vf^i  \rangle } \Big|_{ \langle \vf \rangle =0} ~.
\label{FR1}
\ee
Here the functional derivative 
$\d \G[\langle \vf \rangle, \f ;\c] / \d \langle \vf \rangle $
at $\langle \vf \rangle = 0$ can be related to 
$\d {\bf \G}[\f;\c] / \d \f$  
with the aid of the identity (see \cite{KM} 
and the last reference in \cite{tH} for a derivation)
\bea
\d \f^i \, \frac{\d {\bf \G}[\f;\c]}{\d \f^i}
&=& \Big\{ \d \f^i + 
\langle R^i{}_\a [\f +\vf] \,
(F^{-1} [\vf,\f])^\a{}_\b  \, \D \c^\b [\vf, \f] \rangle 
\Big\}\,
\frac{\d \G[\langle \vf \rangle, \f ;\c]} 
{\d \langle \vf^i  \rangle } \Big|_{ \langle \vf \rangle =0}~, \non \\
\D \c^\a [\vf, \f] &=& \c^\a [\vf -\d \f, \f +\d \f] 
-\c^\a [\vf, \f] ~,
\label{FR2}
\eea
with $\d \f^i$ an arbitrary variation of the background fields.
Eqs. (\ref{FR1}) and (\ref{FR2}) show that any change of the gauge 
conditions is equivalent to a nonlocal field redefinition 
in the effective action.

\sect{Quantum deformation of rigid symmetries}

In this section, we briefly provide an overview of rigid 
anomaly-free symmetries 
of the effective action, see \cite{KM} for more details,
and then give a general discussion of 
the quantum deformation of the conformal symmetry 
in $\cN=4$ SYM theory. 

Let the classical action be invariant,
$S[ \F + \e \, \O[\F] ] = S[\F]$,
with respect to a rigid transformation\footnote{It is sufficient
for the purposes of the present paper to consider linear rigid
classical symmetries only.}
\be
\d \F^i = \e \, \O^i[\F] ~, 
\qquad \O^i [\F] = \O^i{}_j \, \F^j~,
\label{glob}
\ee
where $\O^i{}_j$ is a given field-independent operator
and $\e$ an arbitrary infinitesimal constant parameter.
We will assume several additional properties of the
structure of the gauge and global transformations:
\bea
\O^i{}_{,i} [\F] & = & 0 ~, \label{div} \\
R^i{}_{\a,j} [\F] \,\O^j [\F] -
\O^i{}_{,j}[\F] \, R^j{}_\a [\F]
&=& R^i{}_{\b} [\F] \,f^\b{}_\a [\F] ~, \label{algebra} \\
f^\a{}_\a [\F] &=& 0~. \label{contr}
\eea
Eq. (\ref{div}) ensures that the transformation
$\F^i \to \F^i +\e \, \O^i[\F]$ is unimodular.
Eq. (\ref{algebra}) implies that the commutator of a gauge
transformation with a global symmetry transformation is a gauge
transformation.

At the quantum level, one has to specify  some set of gauge 
conditions, $\c^\a [\F]$, and a  gauge fixing functional, 
$S_{\rm GF}[\c]$. An additional assumption 
we make concerns the behaviour of
the gauge conditions under the symmetry transformations. We assume
\be
\d_\e \c^\a [\F] \equiv
\e \, \c^\a{}_{,i} [\F] \, \O^i[\F]
=\e \,\Big( \L^\a{}_\b \, \c^\b [\F] + \r^\a [\F] \Big)~, 
\label{gauge-cond-tran}
\ee
with $\L^\a{}_\b$ a field independent operator.
It will also be assumed that the homogeneous term 
on the right hand side
leaves $S_{\rm GF}[\c] $ invariant,
$ S_{\rm GF}[\c^\a + \e \, \L^\a{}_\b \, \c^\b]
= S_{\rm GF}[\c^\a ]$.
Under all the above assumptions, the symmetry of 
the quantum theory 
can be shown \cite{KM} to be governed by 
the Ward identity 
\be
\G_{,i} [\f;\c] \; \O^i [\f] 
~=~  \G_{,i} [\f;\c] \; \langle R^i{}_\a [\F] \,
(F^{-1} [\F] )^\a{}_\b \,
\r^\b [\F ] \rangle ~,
\label{ST2}
\ee
which is nothing but the condition of invariance
under quantum mechanically corrected symmetry transformations.

It is easy to generalize the Ward identity (\ref{ST2})
to the background field formulation. 
Assuming that the background covariant gauge conditions 
$\c^\a [\vf, \f]$ transform by  the rule
\be
\d_\e \c^\a [\vf, \f] 
= \c^\a [\vf + \e \,\O[\vf] , \f+ \e\,\O[\f] ] -  \c^\a [\vf, \f]
=\e \,\Big( \L^\a{}_\b \, \c^\b [\vf, \f] + \r^\a [\vf, \f] \Big)~, 
\label{gauge-cond-tran-2}
\ee
and that the gauge fixing functional is invariant under 
(\ref{gauge-cond-tran-2}) with $\r^\a [\vf, \f] =0$, one gets
the following Ward identity
\be
\O^i [\f] \, \frac{\d {\bf \G}[\f;\c]}{\d \f^i}
= \langle R^i{}_\a [\f +\vf] \,
(F^{-1} [\vf,\f])^\a{}_\b  \, \r^\b [\vf, \f] \rangle \;
\frac{\d \G[\langle \vf \rangle, \f ;\c]} 
{\d \langle \vf^i  \rangle } \Big|_{ \langle \vf \rangle =0} ~,
\label{def-final}
\ee
which has to be treated in conjunction with (\ref{FR2})
to express the functional derivative 
$\d \G[\langle \vf \rangle, \f ;\c] / \d \langle \vf \rangle $
at $\langle \vf \rangle = 0$ via 
$\d {\bf \G}[\f;\c] / \d \f$.  

Eq. (\ref{def-final}) determines the true
rigid symmetry of the effective action. In general, the correlation
function in the right hand side of (\ref{def-final}) is a nonlocal 
functional of  the fields. If the effective action is 
computed in the framework of the derivative expansion, 
this correlation function can be represented 
as an infinite series 
of local terms with increasing number of derivatives.
As a rule, this series cannot be truncated at a given order
without spoiling the algebra of rigid symmetry transformations.   
As has been shown before, any change of gauge conditions
is equivalent to a special nonlocal field redefinition in 
the effective action, the latter inducing a modification 
to the structure of symmetry transformations. 
Such nonlocal field redefinitions
will always lead to a re-organization of the derivative expansion 
of the effective action. The freedom to choose gauge conditions 
can therefore be used to seek
field redefinitions which are best adapted to the expression of
symmetries in the context of the derivative expansion of the
effective action. In the case of $\cN = 4$ SYM, for example, use of
't Hooft gauge makes it possible to consistently truncate the
one-loop deformation of conformal symmetry to the terms of first
order in derivatives, as given in eq. (\ref{L=1}), 
in that the corresponding transformations 
(\ref{A-def}) and (\ref{Y-def}) form a field
realization of the conformal algebra without the need to include the
higher derivative terms in the modified quantum symmetry.

In principle, there is nothing wrong with the existence 
of a nonlocal gauge  which, 
when implemented instead of 't Hooft gauge, 
would effectively convert the AdS conformal transformations  
(\ref{A-def}) and (\ref{Y-def}) 
to the unmodified form (\ref{conf}); 
there is, however, one major problem with nonlocal gauges --
it is not known how to consistently define quantum theory.
We would like to sketch a field redefinition induced
by such gauge conditions. 
Let $\bY_\m$ and $\bA_m$ be primary conformal scalar and vector 
fields of canonical mass dimension,  $d_\bY = d_\bA =1$, 
such that $\bY^2 =\bY_\m \bY_\m \neq 0$. 
Using their conformal transformation  laws, 
given in eq.  (\ref{conf}), 
one readily derives the conformal 
variations of their descendants
(including the field strength 
$\bF_{mn} = \pa_m \bA_n - \pa_n \bA_m$):
\bea
- \d \,(\pa_m \bY_\m) &=& (v +2\s)\, \pa_m \bY_\m 
+ \o_m{}^n \,\pa_n \bY_\m + (\pa_m \s) \,\bY_\m ~, \non \\
%- \d \,(\pa^m \pa_m \bY )&=& (v 
%+ 3\s )\,\pa^m \pa_m \bY ~, \label{desc}\\
-\d \, \bF_{mn} &=& (v +2\s)\, \bF_{mn}
+ \o_m{}^p \,\bF_{pn} +\o_n{}^p \,\bF_{mp} ~.
\label{desc}
\eea
Now consider  replacing the variables $\bY_\m$ and $\bA_m$ by new ones,
$Y_\m$ and $A_m$,  which (i) have the same canonical dimension;
(ii) are given by  series in powers of derivatives of 
$\bY_\m$ and $\bA_m$; (iii) possess the $AdS_5 \times S^5$ 
transformations (\ref{A-def}) and (\ref{Y-def}). 
To the leading order in derivatives of the fields, 
the new variables are\footnote{There is also freedom 
to add terms containing factors of the free equations of motion, 
$\Box \bY_\m$ and $\pa^n \bF_{mn}$, which transform covariantly 
under the conformal group.}
\bea
Y_\m &=& \bY_\m - {1\over 4} R^4 \left\{
\frac{ (\pa^n \bY_\m) \, \pa_n\bY^2 }{\bY^4}
- \bY_\m \, \frac{  (\pa^n \bY_\n) \,\pa_n \bY_\n}{\bY^4} 
\right\} ~+~O(\pa^3)~, \non \\ 
A_m &=& \bA_m +{1\over 4} R^4  
\left\{  \frac{ \bF_{mn} \,\pa^n\bY^2 }{\bY^4} 
- \pa_m \Big( \frac{ \bA_{n} \,\pa^n\bY^2 }{\bY^4} \Big) \right\} 
~+~O(\pa^3)~,
\label{Min-->AdS}
\eea
as can be checked with the use of (\ref{desc}).
It is not difficult to convince oneself that such 
a field redefinition  can be reconstructed order by order 
in the derivative expansion.
Making the field redefinition (\ref{Min-->AdS}) 
in the D3-brane action (\ref{d3brane}), one ends up with a higher 
derivative action which is invariant under the classical 
transformations (\ref{conf}).

\sect{Deformation of conformal symmetry in \mbox{$R_\x$} gauge}

Here we illustrate the general analysis given in the
preceding sections by explicit calculations of the quantum 
deformation of conformal symmetry in the so-called $R_\x$ gauge. 
We are interested in the bosonic sector
of the $\cN=4$ super Yang-Mills theory described by
fields $\F^i = \{A_m (x), Y_\m (x)\} $, 
where $m=0,1,2,3$ and $\m=1,\ldots, 6$.
The classical action is
\be
S [A,Y] = -\frac{1}{4g^2}\int {\rm d}^{4} x \; {\rm tr} \Big(
F^{mn} F_{mn} + 2 D^m Y_\m D_m Y_\m
- [Y_\m, Y_\n]\, [Y_\m, Y_\n] \Big)~,
\label{action}
\ee
with $D_m = \pa_m +{\rm i} A_m$, and is invariant under standard
gauge transformations
\be
\d A_m = - D_m \t = - \pa_m \t -{\rm i}\,[A_m, \t]~,
\qquad \d Y_\m = {\rm i}\, [\t, Y_\m]~.
\ee
The theory is quantized in the background field approach, i.e.
by splitting the dynamical variables
$\F^i $ into the sum of background
fields $\f^i = \{\cA_m (x), \cY_\m(x) \}$ and quantum
fields $\vf^i = \{a_m(x), y_\m(x)\}$. In $R_\x$ gauge,  
the gauge conditions are
\be
\c^{(\x)}  = \frac{1}{\sqrt{\x}}\, \cD^m a_m + 
{\rm i} \,\sqrt{\x}\,[\cY_\m , y_\m ] ~,
\label{Rxgauge}
\ee
where $\cD_m$ are the background covariant derivatives, 
and $\x$ the gauge fixing parameter.
The gauge fixing functional, $S_{\rm GF}$, is the same 
as in \cite{KM}
\be
S_{\rm GF} [\c^{(\x)}] = -\frac{1}{ 2g^2} \int {\rm d}^4 x \;
{\rm tr} \, (\c^{(\x)})^2 ~.
\ee
The choice $\x=1$ corresponds to 't Hooft gauge
implemented in our previous work \cite{KM}.

Under the combined conformal transformation (\ref{conf})
of the background and quantum fields, $\c^{(\x)}$ changes as follows 
\be
\d_{\rm c} \c^{(\x)} =  - (v +2 \s) \, \c^{(\x)}  
+\frac{2}{\sqrt{\x}} \, (\pa^m \s ) \, a_m~,
\label{gauge-cond-conf-tran}
\ee
and this transformation law is clearly of the form 
(\ref{gauge-cond-tran-2}). The inhomogeneous term in 
(\ref{gauge-cond-conf-tran}) is the source of  a quantum
modification to the conformal Ward identity, 
which can be computed in $R_{\xi}$ gauge by extension of  
the 't Hooft gauge calculation in \cite{KM}.
Choosing a $U(1)$ background which spontaneously breaks 
the gauge group $SU(N+1)$ to
$SU(N)\times U(1)$, and retaining only terms of first order in derivatives,
the one-loop modification to the conformal transformations is:
\bea 
\d^{(\x)} A_m
&=& - \, \frac{N g^2}{4 \pi^2} \, 
\frac{ (\partial^n \sigma) F_{mn}  } {Y^2}~, \label{Rx1}\\
\delta^{(\x)} Y_\m &=&  
\frac{N g^2}{4 \pi^2} \, 
\frac{ (\partial^n \sigma) (\partial_n Y_\m)}{Y^2}
\Big[  
\frac{\ln \x}{(\x-1)^2} 
+ \frac{\ln \x}{(\x-1)}
- \frac{1}{(\x-1)} +\hf \Big] \non \\
&-& \frac{N g^2}{8 \pi^2} \, 
\frac{ (\partial^n \sigma) (\partial_n Y^2) Y_\m}{Y^4}
\Big[ {5 \over 2} \,\frac{\ln \x}{(\x-1)^2} 
+ \frac{\ln \x}{(\x-1)} 
- {5\over 2} \,\frac{1}{(\x-1)} +{1 \over 4} \Big] ~.
\label{Rx2}
\eea
As can be seen, the gauge field transformation 
is the same as in 't Hooft gauge. 
In relation to the scalar transformation  (\ref{Rx2}), 
the second square bracket in $\delta^{(\x)} Y_\m $
vanishes in the limit $\x \to 1$ ('t Hooft gauge), 
and the first square bracket gives 1, thus yielding 
the 't Hooft gauge result. The transformations (\ref{Rx1})  and 
(\ref{Rx2}) do {\it not} realize the conformal algebra
for $\x \neq 1$. This means that, when computing 
the quantum modification to the conformal transformations in $R_\x$ gauge, 
we have to take into account the terms of second and higher orders
in derivatives of the fields. 

Let us analyse a special case, $\x=1+\ve$, with an  infinitesimal 
parameter  $\ve$. 
Then eq. (\ref{Rx2}) reduces to 
\bea
\delta^{(1+\ve)} Y_\m &=&  
\frac{N g^2}{4 \pi^2} \, 
\frac{ (\partial^n \sigma) (\partial_n Y_\m)}{Y^2}
\Big[ 1 - {\ve \over 6} \Big] ~-~ \frac{N g^2}{8 \pi^2} \, 
\frac{ (\partial^n \sigma) (\partial_n Y^2) Y_\m}{Y^4}
\Big[ {\ve \over3 }\Big]~.
\eea
In accordance with the previous discussion,
there should exist a field redefinition 
relating the fields $Y_\m$ in $R_{1+ \ve}$ gauge to those, 
$\tilde{Y}_\m$, in 't Hooft gauge. It is
\be 
Y_\m = \tilde{Y}_\m + \ve \, \frac{N g^2}{48 \pi^2} \,
\Big\{  \frac{ (\pa^n \tilde{Y}_\m) (\pa_n \tilde{Y}^2) }
{\tilde{Y}^4}
~+ ~
\frac{ (\pa^n \tilde{Y}_\n) (\pa_n \tilde{Y}_\n) }
{\tilde{Y}^4} \, \tilde{Y}_\m \Big\}  ~+~O(\pa^3)~.
\ee

\sect{Conclusion}

In this paper, we have addressed several issues 
related to the gauge dependence of the quantum deformation 
of the conformal symmetry in $\cN=4$ SYM.
In section 3, we have extended some well known results on the
gauge dependence of the effective action to the case of background
field quantization, and the result (\ref{FR1}) demonstrates that a change
of background covariant  gauge conditions
is equivalent to a nonlocal field redefinition.
In section 4, we derived the quantum-corrected Ward identity
(\ref{def-final}) in the background field approach. 
The explicit one-loop calculations in $R_{\xi}$ gauge of the quantum
modifications to conformal symmetry in $\cN=4$ SYM in section 5
highlight the very special nature of 't Hooft gauge for this theory.
The modified conformal transformations 
(\ref{Rx1}) and (\ref{Rx2})  do not form a
closed algebra when truncated at first derivative order in the
derivative expansion, except in the case $\xi = 1,$ namely 't Hooft
gauge. Only in 't Hooft gauge, 
the truncated transformations coincide with
the symmetry transformations (\ref{A-def}) and (\ref{Y-def})
of a D3-brane embedded in $AdS_5 \times S^5.$ 
Also striking is the fact that the quantum
modification of the transformation of the gauge field is not modified by
moving out of 't Hooft gauge to $R_{\xi}$ gauge.
As was demonstrated in [2], 't Hooft gauge retains some memory of the
origin of classical $\cN = 4$ supersymmetric Yang-Mills theory  as a
dimensional reduction of ten-dimensional supersymmetric
Yang-Mills theory. It would be of interest to determine if this is of
any significance in relation to these observations.

The phenomenon of quantum deformation of rigid symmetries is quite
general. Apart from conformal symmetry in $\cN=4$ super Yang-Mills 
theory, it is also worth mentioning here nice results on 
generalized conformal symmetry in D-brane matrix models 
\cite{JKY2,HM} and supersymmetry in Matrix theory
\cite{KaMu}.

We believe that the deformed conformal invariance
in the $\cN=4$ super Yang-Mills theory should be
crucial
%, along with the requirement
%of nonlinear self-duality \cite{GKPR,KT},
for a better understanding  of numerous non-renormalization
theorems which are predicted by the AdS/CFT conjecture
and relate to the explicit structure of the
low energy effective action in $\cN=4$ super Yang-Mills
theory (see \cite{CT,BKT,BPT} for a more detailed discussion
and  additional references).

While this work was in the process of being written up, 
a paper appeared on the hep-th archive \cite{BIK} where 
the techniques of nonlinear realizations were used 
to derive a field redefinition similar to the one 
introduced at the end of section 4.

\vskip.5cm

\noindent
{\bf Acknowledgements.}
Discussions with  N. Dragon, O. Lechtenfeld, D. L\"ust, B. Zupnik,  
and especially with J. Erdmenger and S. Theisen,
are gratefully acknowledged. 
This work is partially supported by a University of Western 
Australia Small Grant and an ARC Discovery Grant.
One of us (SMK) is thankful to N. Dragon and S. Theisen 
for kind hospitality  extended to him 
at the Max Planck Institute for Gravitational 
Physics (Albert Einstein Institute), Golm 
and at the Institute for Theoretical Physics, Uni-Hannover.
The work of SMK has been supported in part by the Alexander 
von Humboldt Foundation, the Max Planck Society and the
German National Science Foundation.

\end{document}